\documentclass[aps,twocolumn,nofootinbib, superscriptaddress, floatfix]{revtex4-1}

\usepackage{multirow}
\usepackage{booktabs}    
\usepackage{placeins}
\usepackage{soul}
\usepackage[usenames,dvipsnames,svgnames]{xcolor}
\definecolor{redd}{rgb}{0.8, 0.1,0.2}
\definecolor{navy}{rgb}{0.05, 0.23,0.75}
\usepackage[colorlinks]{hyperref}
\hypersetup{
     colorlinks   = true,
     citecolor    = navy,
	linkcolor = redd,
	urlcolor=navy,
	anchorcolor=blue
}
\usepackage{bbm}
\usepackage{amsfonts}
\usepackage{amsmath}
\usepackage{mathrsfs}
\usepackage{soul}

\usepackage{amsmath,amssymb,amsbsy,bm}
\usepackage{graphicx}
\usepackage{wrapfig}                 
\usepackage{url}
\usepackage{hyperref}
\usepackage{float}
\usepackage{pstricks}
\usepackage{color}
\usepackage{multirow}
\usepackage{lipsum}
\usepackage{enumitem}
\usepackage{ctable}
\newcolumntype{L}{>{\centering\arraybackslash}m{1.5cm}}

\usepackage{titlesec}
\titleformat{\section}[runin]{\normalfont \bfseries}{\thesection}{1em}{}
\titleformat{\subsection}[runin]{\normalfont \bfseries}{\thesubsection}{1em}{}

\usepackage{enumitem}

\newcommand{\bear}{\begin{array}}
\newcommand {\eear}{\end{array}}

\newcommand{\beq}{\begin {equation}}
\newcommand{\eeq}{\end   {equation}}
\newcommand{\bea}{\begin {eqnarray}}
\newcommand{\eea}{\end   {eqnarray}}
\newcommand{\baa}{\begin {array}   }
\newcommand{\eaa}{\end   {array}   }
\newcommand{\bit}{\begin {itemize} }
\newcommand{\eit}{\end   {itemize} }
\newcommand{\be }{\begin {equation}}
\newcommand{\ee }{\end   {equation}}
\newcommand{\nn }{\nonumber        }

\def\bea{\begin{eqnarray}}
\def\eea{\end{eqnarray}}

\newcommand{\bef}{\begin{figure}}
\newcommand {\eef}{\end{figure}}
\newcommand{\bec}{\begin{center}}
\newcommand {\eec}{\end{center}}

\definecolor{cerulean}{rgb}{0., 0.62,0.7}

\newcommand{\twiddle}{{\raise.17ex\hbox{$\scriptstyle\sim$}}}

\begin{document}

\title{\Large\color{navy} Minimal Neutral Naturalness Model}

\author{Ling-Xiao Xu}
\email{lingxiaoxu@pku.edu.cn}
\affiliation{Department of Physics and State Key Laboratory of Nuclear Physics and Technology, \\Peking University, Beijing 100871, China}

\author{Jiang-Hao Yu}
\email{jhyu@itp.ac.cn}
\affiliation{CAS Key Laboratory of Theoretical Physics, Institute of Theoretical Physics, Chinese Academy of Sciences, Beijing 100190, P. R. China}
\affiliation{School of Physical Sciences, University of Chinese Academy of Sciences, No.19A Yuquan Road, Beijing 100049, P.R. China}

\author{Shou-hua Zhu}
\email{shzhu@pku.edu.cn}
\affiliation{Department of Physics and State Key Laboratory of Nuclear Physics and Technology, \\Peking University, Beijing 100871, China}
\affiliation{Collaborative Innovation Center of Quantum Matter, Beijing, 100871, China}
\affiliation{Center for High Energy Physics, Peking University, Beijing 100871, China}

\begin{abstract}

We build a minimal neutral naturalness model in which the top partners are not charged under QCD, with a pseudo Goldstone Higgs arising from $SO(5)/SO(4)$ breaking. 
The color-neutral top partners generate the Higgs potential radiatively without quadratic divergence. 
The misalignment between the electroweak scale and global symmetry breaking scale is naturally obtained from suppression of the Higgs quadratic term, due to cancellation between singlet and doublet top partner contributions.
This model can be embedded into ultraviolet holographic setup in composite Higgs framework, which even realizes finite Higgs potential.

\end{abstract}

\maketitle

\section{Introduction}
\label{sec:intro}

The hierarchy problem remains as one of the unsolved puzzles in the Standard Model (SM), i.e. the Higgs mass is sensitive to the Planck scale through quantum effects. Symmetry for the Higgs boson, such as supersymmetry and shift symmetry, is typically introduced to relate the top quark with top partner, which lowers sensitivity of the ultraviolet (UV) scale in the Higgs mass down to the top partner scale. However, current experimental searches of the colored top partners has already set the lower bound of their masses around $1$ TeV~\cite{Aaboud:2017ayj,ATLAS:2018iwl,Sirunyan:2016jpr,Khachatryan:2017rhw}, which lead to the little hierarchy problem~\cite{Barbieri:2000gf}.

One novel solution to the little hierarchy problem is neutral naturalness scenario~\cite{Craig:2015pha, Craig:2014aea, Chacko:2005pe,Burdman:2006tz,Cai:2008au,Cohen:2018mgv}, in which top partners are not charged under the SM color group, i.e. twin Higgs~\cite{Chacko:2005pe}, folded supersymmetry~\cite{Burdman:2006tz}, quirky little Higgs~\cite{Cai:2008au}, hyperbolic Higgs~\cite{Cohen:2018mgv}. The current search limit on masses of the colorless top partners is still below TeV~\cite{Cheng:2015buv,Chacko:2015fbc}, which softens the little hierarchy problem.
Twin Higgs is the first example of neutral naturalness, in which the Higgs boson is identified as a pseudo Nambu-Goldstone boson (PNGB) due to an accidental $Z_2$ between the SM and its twin copy. 
Although this idea is conceptually simple, it introduces many particles in the hidden sector, i.e. mirror $W^\prime$, $Z^\prime$, $\gamma^\prime$ and whole generation of chiral fermions for anomaly cancellation.
As a result, this setup for hidden sector suffers from cosmological constraints due to the presence of hidden neutrinos and hidden photons~\cite{Craig:2016lyx,Chacko:2016hvu,Csaki:2017spo,Chacko:2018vss}.
It is well-motivated to find alternative constructions for more minimal hidden sector: mirror copies of the SM gauge bosons are not necessary, vector-like fermions instead of chiral ones are introduced, i.e., no need for the whole generation of fermions.

In this paper, we present concrete neutral naturalness models with the minimal hidden sector. 
We only introduce the dark $SU(3)_c^\prime$ gauge symmetry in the hidden sector and adopt the \emph{minimal} $SO(5)/SO(4)$ coset~\cite{Agashe:2004rs} for the PNGB Higgs. We also introduce \emph{minimal} numbers of elementary vector-like fermions: one singlet and one doublet of $SU(2)_L$, for cancelling quadratic divergence from the top quark contribution, and realizing vacuum misalignment~\cite{Kaplan:1983fs} between the electroweak scale $ v$ and the global symmetry breaking scale $f$, i.e., $v\ll f$.
We denote our model as \emph{the minimal neutral naturalness model} (MNNM). Our setup provides a new way to generate the electroweak scale via fully radiative symmetry breaking in the Higgs potential. The potential of the PNGB Higgs can generally be parametrized~\cite{Marzocca:2012zn,Bellazzini:2014yua} as
\bea
V(h)\simeq-\gamma \sin^2\left(\frac{h}{f}\right)+\beta \sin^4\left(\frac{h}{f}\right),
\label{gen_po}
\eea
where $\gamma$ and $\beta$ denote radiative corrections from gauge boson and fermion contributions, and the electroweak scale is obtained by $\frac{v^2}{f^2}= \frac{\gamma}{2\beta}$. 
Radiative Higgs potential typically gives $\gamma \simeq \beta$ and thus $v\simeq f$.
To realize vacuum misalignment $v\ll f$,
one needs to either suppress the value of $\gamma$ or increase $\beta$. In twin Higgs, $\gamma$ can be suppressed by cancellation between fermion and gauge boson contributions~\cite{Yu:2016bku, Yu:2016swa}. In this work, we show for the first time this suppression of $\gamma$ can happen due to cancellation between fermion contributions, i.e., the SM top quark and the color-neutral top partners.

To have our setup valid at UV scale, we extend MNNM by including composite states in the holographic framework~\cite{ArkaniHamed:2000ds}. Following spirit of composite Higgs models~\cite{Contino:2010rs,Panico:2015jxa, Agashe:2004rs,Geller:2014kta,Barbieri:2015lqa,Low:2015nqa,Csaki:2017cep,Serra:2017poj,Csaki:2017jby,Dillon:2018wye}, we present a holographic MNNM and its deconstructed version.
This brings finiteness of the Higgs potential, i.e., not sensitive to  the UV cutoff. After integrating out composite states, we recover the MNNM spectrum.

\section{The Model}
\label{sec:model}
Let us first introduce the field content of the hidden sector. 
The SM gauge symmetry is extended to $SU(3)^\prime_c \times SU(3)_{c} \times SU(2)_L \times U(1)_Y$ where $SU(3)^\prime_c$ is an unbroken dark color gauge group.
We introduce two vector-like fermions $\widetilde{q}\equiv (\widetilde{t},\widetilde{b})^T$ and $\widetilde{T}$, which are QCD neutral but carry the dark QCD charge,
\bea
	\widetilde{q} \sim (3, 1, 2)_Y, \ \ \  \widetilde{T} \sim (3, 1, 1)_Y,
\eea
under $SU(3)^\prime_c \times SU(3)_{c} \times SU(2)_L \times U(1)_Y$, with $Y$ arbitrarily chosen.
The vector-like fermions play the role of top partners with global $SU(6)$ or $Z_2$ parity.
Typically introducing only the singlet $\widetilde{T}$ is enough for quadratic divergence cancellation. However, to obtain the realistic Higgs potential with vacuum misalignment, additional matter content is needed. Here  the additional doublet $\widetilde{q}$ is introduced for the purpose.

All the fermion contents are embedded into representations of global symmetry, in which the Higgs is a PNGB from global symmetry breaking.
We adopt the minimal coset $SO(5)/SO(4)$~\cite{Agashe:2004rs} incorporating the custodial symmetry in the Higgs sector, but do not include any compositeness from strong dynamics.
The Higgs boson is represented by the nonlinear sigma field in unitary gauge as 
\bea
\Sigma=f\ (0,0,0,s_h,c_h)^T\ ,
\eea
where $s_h\equiv \sin(h/f)$, $c_h\equiv \cos(h/f)$.
The SM $SU(2)_L\times U(1)_Y$ symmetry is embedded in $SO(4)\times U(1)_X\cong SU(2)_L\times SU(2)_R\times U(1)_X$ with the hyper-charge
$Y=X+T^3_R$.
The additional $U(1)_X$ is needed to obtain the correct hyper-charge for the SM quarks.
The SM doublet $q_L=(t_L, b_L)^T$ and singlet $t_R$ are embedded into $5$-plet and singlet of the $SO(5)$ respectively 
\bea
Q_L&=\frac{1}{\sqrt{2}}
\left(
\baa{c}
b_L\\-ib_L\\t_L\\it_L\\0 
\eaa
\right)\subset {\bf{5}},\quad \quad t_R \subset {\bf{1}}.
\label{smfermion}
\eea
with quantum number $X=2/3$.
In the hidden sector, the vector-like fermions $\widetilde{q}_{L,R}$ and $\widetilde{T}_{L,R}$ are embedded as follows
\bea
\widetilde{Q}_L=\frac{1}{\sqrt{2}}
\left(
\baa{c}
\widetilde{b}_L\\-i\widetilde{b}_L\\\widetilde{t}_L\\i\widetilde{t}_L\\\sqrt{2}\widetilde{T}_L 
\eaa
\right)\subset {\bf{5}},\ 
\widetilde{Q}_R&=\frac{1}{\sqrt{2}}
\left(
\baa{c}
\widetilde{b}_R\\-i\widetilde{b}_R\\\widetilde{t}_R\\i\widetilde{t}_R\\0 
\eaa
\right)\subset {\bf{5}},
\ \nn\\
\widetilde{T}_R \subset {\bf{1}},
\label{hiddenfermion}
\eea
with arbitrary $U(1)_X$ charge.

After embedding all the above elementary fermions in $SO(5)$ representations, we write down the following Lagrangian for the top Yukawa sector
\bea
-\mathcal{L}_{\text{top}}=\ y\bar{Q}_L\Sigma t_R+\tilde{y}\bar{\widetilde{Q}}_L\Sigma \widetilde{T}_R-m_{\widetilde{q}}\bar{\widetilde{Q}}_L\widetilde{Q}_R+\textnormal{h.c.}\ .\ 
\label{yukawa}
\eea
There is no mixing between the SM top quark and the top partners as they carry different $SU(3)^\prime_c \times SU(3)_{c}$ charges.
The mass matrix of the hidden top partners reads
\bea
\mathcal{L}_{\text{mass}}\supset (\bar{\widetilde{t}}_L, \bar{\widetilde{T}}_L)
\left(
\baa{cc}
m_{\widetilde{q}}& \frac{i\widetilde{y}f}{\sqrt{2}}s_h\\ 
0 & -\widetilde{y}fc_h
\eaa
\right)
\left(
\baa{c}
\widetilde{t}_R\\
\widetilde{T}_R
\eaa
\right)+\text{h.c.}\ .
\eea
The doublet-singlet fermion mixing is approximately proportional to $\widetilde{y}fs_h/m_{\widetilde{q}}$. Taking $m_{\widetilde{q}}\to \infty$ gives rise to the decoupling limit of the doublet fermion. 

In order to cancel quadratic divergence from the top quark loop, additional symmetry needs to be imposed in the top Yukawa terms in Eq.~\ref{yukawa}.
Here we adopt the global $SU(6)_c$ symmetry in the Yukawa sector with $SU(3)_c$ and $SU(3)_c^\prime$ gauged: $SU(3)_c\times SU(3)_c^\prime\subset SU(6)_c$. Under the global symmetry $SO(5)\times SU(6)_c$, we define the bi-fundamental fermions based on Eqs.~\ref{smfermion} and~\ref{hiddenfermion}
\bea
\mathcal{Q}_L=(Q_L, \widetilde{Q}_L), \ \ \mathcal{T}_R=(t_R, \widetilde{T}_R).
\eea
The Yukawa terms in Eq.~\ref{yukawa} are rewritten as
\bea
\mathcal{L}_{\text{top}}\supset y\bar{\mathcal{Q}}_L\Sigma \mathcal{T}_R+\textnormal{h.c.},\quad \text{with}\quad y=\widetilde{y}.
\eea
Alternatively, introducing a $Z_2$ symmetry in Eq.~\ref{yukawa} 
\bea
Z_2:\quad y\bar{Q}_L\Sigma t_R \ \longleftrightarrow \ \tilde{y}\bar{\widetilde{Q}}_L\Sigma \widetilde{T}_R
\eea
also causes $y=\widetilde{y}$, which induces quadratic divergence cancellation. Different from the twin Higgs $Z_2$ parity~\cite{Chacko:2005pe}, this $Z_2$ symmetry introduced is explicitly broken, thus the Higgs potential is not symmetric under $s_h\leftrightarrow c_h$ in the model.

\section{Fully Radiative Higgs Potential}

Following Coleman-Weinberg~\cite{Coleman:1973jx}, one obtains the one-loop Higgs potential. Instead of presenting lengthy expressions, we demonstrate the vacuum misalignment is obtained naturally with mass insertion method.

\begin{figure}[!h]
\includegraphics[scale=0.6]{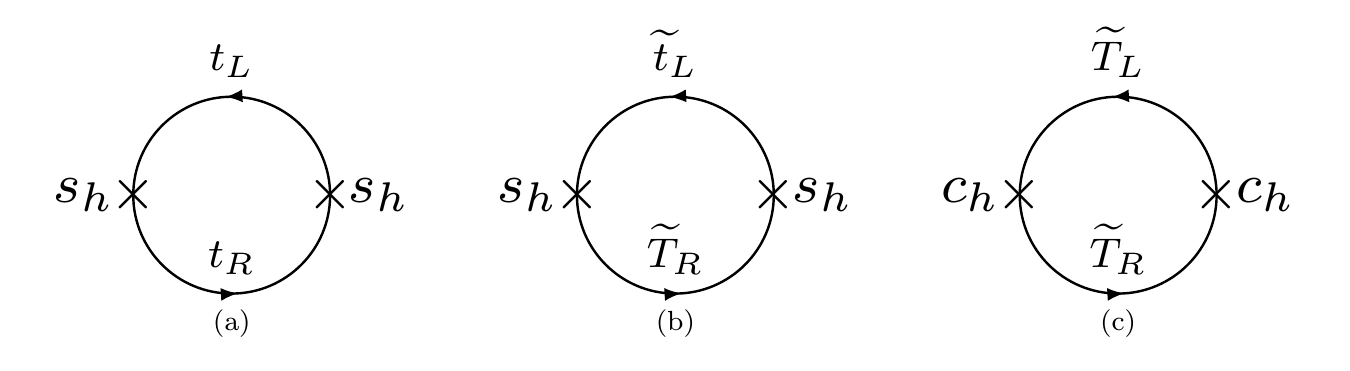}
\caption{Quadratic divergence cancellation between the top quark and the top partners using mass insertion. Each cross ($\times$) denotes a Higgs insertion.}
\label{fig1}
\end{figure}

First of all, let us address quadratic divergence cancellation when additional symmetries, i.e., $SU(6)$ or $Z_2$, impose $y=\widetilde{y}$. According to Fig.~\ref{fig1}, the Higgs quadratic term from each diagram reads
\bea
V(h)\sim\frac{y^2f^2N_c\Lambda^2}{16\pi^2}\left(\frac{1}{2}s_h^2+\frac{1}{2}s_h^2+c_h^2\right),
\label{nat}
\eea
where $N_c=3$ and $\Lambda$ denotes the UV cutoff. The mass parameter $m_{\widetilde{q}}$ is irrelevant to $\Lambda^2$. Even if the doublet fermion is decoupled ($m_{\widetilde{q}}\to\infty$), the quadratic divergence is cancelled by only the singlet fermion, with its kinetic term being renormalized.

\begin{figure}[!h]
\includegraphics[scale=0.6]{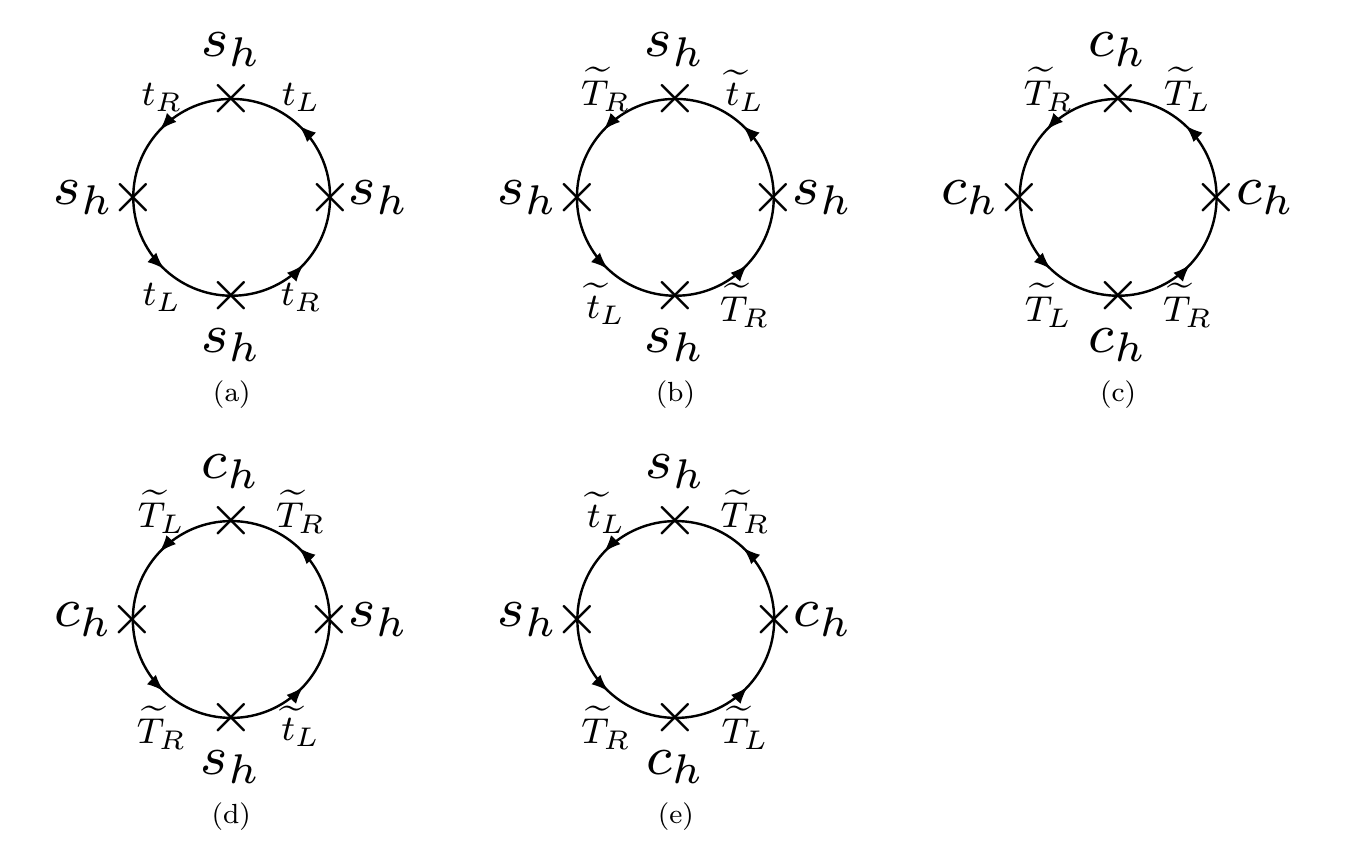}
\caption{The logarithmic divergent Higgs potential induced by only the Yukawa terms in Eq.~\ref{yukawa}. }
\label{fig2}
\end{figure}

Then we investigate the logarithmic divergence of the Higgs potential. We temporally neglect the contribution from the mass term in Eq.~\ref{yukawa}. According to Fig.~\ref{fig2}, the logarithmic divergent part from each diagram reads
\begin{align}
V(h)\sim \ &\frac{y^4f^4N_c\ \text{log}\Lambda^2}{16\pi^2}\left(\frac{1}{4}s_h^4+\frac{1}{4}s_h^4+c_h^4+\frac{1}{2}s_h^2c_h^2+\frac{1}{2}s_h^2c_h^2\right)\nn\\
\sim \ &\frac{y^4f^4N_c\ \text{log}\Lambda^2}{16\pi^2}\left(-s_h^2+\frac{1}{2}s_h^4\right).
\label{pot1}
\end{align}
The electroweak symmetry breaking is triggered by the $c_h^4$ term from the singlet fermion $\widetilde{T}$ loop. Without contribution from the mass term, the electroweak scale is obtained to be $v=f$, which is too large to be compatible with data.

\begin{figure}[!h]
\includegraphics[scale=0.6]{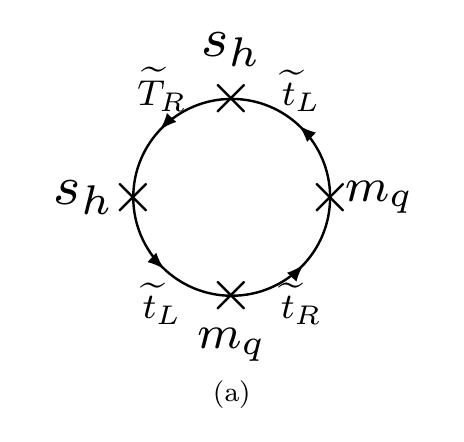}
\caption{The logarithmic divergent Higgs potential induced by the doublet fermion mass term $-m_{\widetilde{q}}\bar{\widetilde{Q}}_L\widetilde{Q}_R$.}
\label{fig3}
\end{figure}

Finally, we include the $\text{log}$-dependent Higgs potential from the mass term $m_{\widetilde{q}}\bar{\widetilde{Q}}_L\widetilde{Q}_R$. According to Fig.~\ref{fig3}, it provides a positive $s_h^2$ contribution 
\bea
V(h)\sim \ &\frac{y^2f^2N_c\ \text{log}\Lambda^2}{16\pi^2}\ m_{\widetilde{q}}^2\ s_h^2.
\label{pot2}
\eea
Combined with the negative $s_h^2$ contribution in Eq.~\ref{pot1}, we obtain the complete $\log$-dependent Higgs potential
\bea
V(h)\sim \ &\frac{y^2f^2N_c\ \text{log}\Lambda^2}{16\pi^2}\left[(m_{\widetilde{q}}^2-y^2f^2)s_h^2+\frac{y^2f^2}{2}s_h^4\right].
\label{log_pot}
\eea
The coefficient of $s_h^2$ term can be much smaller than that of the $s_h^4$ term, due to the cancellation between the Yukawa terms and the doublet fermion mass term. As mentioned in Introduction, unlike cancellation between fermions and gauge bosons in twin Higgs models~\cite{Yu:2016bku, Yu:2016swa}, we provide a novel way to suppress the $s_h^2$ relative to the $s_h^4$ term.

 As a result, we are able to obtain the vacuum misalignment without including contribution from the bosonic sector. The vacuum misalignment is parametrized by
\bea
\xi\equiv\frac{v^2}{f^2}=1-\frac{m_{\widetilde{q}}^2}{y^2f^2}\ .
\eea
Since the top partner mass is naturally at the order of $m_{\widetilde{q}}\sim yf$, $\xi\sim 0.1$ can be realized.
In our numerical study, we further include the finite part of the Higgs potential, which does not change the general result depicted above.

\section{Ultraviolet Realization}
Typically the PNGB Higgs can origin from new strong dynamics above the TeV scale, which introduces compositeness. The strong dynamics can effectively be described with the holographic framework~\cite{ArkaniHamed:2000ds}, more specifically, the composite Higgs, for reviews see~\cite{Contino:2010rs,Panico:2015jxa}.
Introducing composite fermions further gets rid of the $\log\Lambda^2$ dependence in the Higgs potential of Eq.~\ref{log_pot}, with the scale $\Lambda$ being interpreted as composite state masses. In MNNM, the gauge boson loops still encounter $\Lambda^2$ dependence. Introducing composite gauge bosons cancels both the $\Lambda^2$ and $\log\Lambda^2$ dependence.
These render the full Higgs potential finite.

In the extra dimensional setup, the Higgs boson is identified as the zero mode of the fifth dimensional gauge field $A_5(x^\mu,z)$~\cite{ArkaniHamed:2001nc,Contino:2003ve}.
The five-dimensional ($5$D) metric of $AdS_5$ is
$ds^2=\left(\frac{L}{z}\right)^2(\eta_{\mu\nu}dx^\mu dx^\nu-dz^2)$
where the UV and IR branes are localized at $z_{UV}=L_0$ and $z_{IR}=L_1$ respectively. 
The bulk gauge symmetry $SO(5)\times U(1)_X\times SU(3)_c\times SU(3)^\prime_c$ is broken to $SO(4)\times U(1)_X\times SU(3)_c\times SU(3)^\prime_c$ on the IR brane, while is further reduced to $SU(2)_L\times U(1)_Y\times SU(3)_c\times SU(3)^\prime_c$ on the UV brane, with the hyper-charge $Y=X+T^3_R$.

For the fermions in the SM sector, $q_L$ and $t_R$ are respectively identified as the zero modes of the bulk fields $\xi_q$ and $\xi_t$ with corresponding boundary conditions
\begin{align}
\xi_q&=\left[
\baa{cc}
(2,2)^q_L=\left[\baa{c}q^\prime_L(-+)\\q_L(++)\eaa\right]& (2,2)^q_R=\left[\baa{c}q^\prime_R(+-)\\q_R(--)\eaa\right]\\
(1,1)^q_L(-+)&(1,1)^q_R(+-)
\eaa
\right],\nn\\
\xi_t&=\left[\baa{cc}(1,1)^t_L(--)&(1,1)^t_R(++)\eaa\right],
\label{bulksm}
\end{align}
where $(\pm,\pm)$ denote the Neumann ($+$) and Dirichlet ($-$) boundary conditions (B.C.) on the UV and IR branes. 
Note that $\xi_q$ and $\xi_t$ are charged under the $SU(3)_c$ but neutral under the $SU(3)^\prime_c$. The above fermion assignment respects the $SO(5)$ symmetry on the IR brane, which render the Higgs as the exact Goldstone.
We add the IR-brane localized term
\bea
\mathcal{L}\supset \frac{m}{g_5^2}\ \overline{(1,1)^q_L}\ (1,1)^t_R\ (z_{IR}=L_1)+\text{h.c.}\ ,
\eea
where $m$ is dimensionless mass parameter and $g_5$ is $5$D gauge parameter with $\text{Dim}[1/g_5^2]=1$. This term explicitly breaks the $SO(5)$ symmetry on the IR brane and thus generates the top quark mass and the finite Higgs potential.

For the fermions in the hidden sector, the elementary fermion doublet $\widetilde{q}_{R}$ is only localized on the UV brane, while the elementary fermion doublet $\widetilde{q}_{L}$ and singlet $\widetilde{T}_{L,R}$ are embedded in the bulk fermions $\xi_{\widetilde{q}}$ and $\xi_{\widetilde{T}}$ respectively as follows,
\begin{align}
\xi_{\widetilde{q}}&=\left[
\baa{cc}
(2,2)^{\widetilde{q}}_L=\left[\baa{c}\widetilde{q}^\prime_L(-+)\\ \widetilde{q}_L(++)\eaa\right]& (2,2)^{\widetilde{q}}_R=\left[\baa{c}\widetilde{q}^\prime_R(+-)\\ \widetilde{q}_R(--)\eaa\right] \\
(1,1)^{\widetilde{q}}_L(++)&(1,1)^{\widetilde{q}}_R(--)
\eaa
\right],\nn\\
\xi_{\widetilde{T}}&=\left[\baa{cc}(1,1)^{\widetilde{T}}_L(--)&(1,1)^{\widetilde{T}}_R(++)\eaa\right].
\end{align}
The bulk fermions $\xi_{\widetilde{q}}$ and $\xi_{\widetilde{T}}$ are charged under the $SU(3)^\prime_c$ while neutral under the $SU(3)_c$. Similarly, the IR-brane term, which breaks the fermionic $SO(5)$ symmetry on the IR brane, is
\bea
\mathcal{L}\supset\frac{\widetilde{m}}{g_5^2}\ \overline{(1,1)^{\widetilde{q}}_L}\ (1,1)^{\widetilde{T}}_R\ (z_{IR}=L_1)+\text{h.c.}\ .
\eea
The dimensionless mass parameter $\widetilde{m}$ induces additional contribution to the finite Higgs potential and the hidden fermion masses.
According to our UV brane assignment, we introduce the UV-brane localized mass term for the doublet fermion
\bea
\mathcal{L}\supset-\frac{\widetilde{m}_q}{g_5^2}\ \overline{\widetilde{q}}_{R}\ \widetilde{q}_L(++)\ (z_{UV}=L_0)+\text{h.c.}\ ,
\eea
which respects the gauge symmetry on the UV brane.

\begin{figure}[!h]
\includegraphics[scale=0.23]{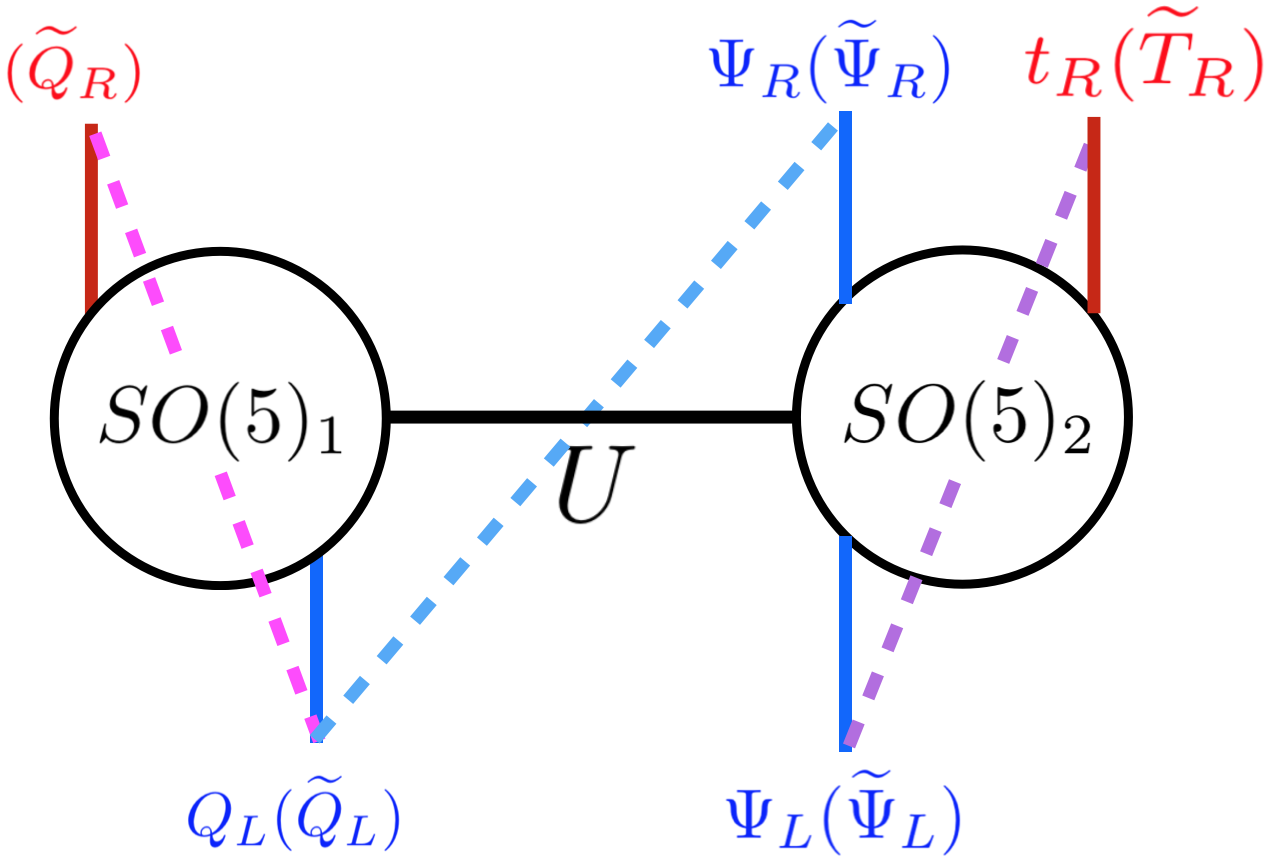}
\caption{The two-site moose diagram and fermion assignment for composite neutral naturalness. Each circle denotes a site and the link in between denotes the Goldstone matrix.}
\label{fig4}
\end{figure}

Following dimensional deconstruction~\cite{ArkaniHamed:2001ca}, the above extra dimensional setup can effectively be described by multi-site moose models~\cite{Georgi:1985hf,Cheng:2006ht}. The minimal moose setup is the two-site model based on the $SO(5)_1\times SO(5)_2/SO(5)_V$ coset~\cite{Foadi:2010bu,Panico:2011pw} with $SO(4)_2$ gauged on the $2$-site. The $\Lambda^2$ dependence from the SM gauge bosons is cancelled by the composite $\rho$ mesons introduced by gauging $SO(4)_2$. This can be viewed as the extension of the $SO(5)/SO(4)$ coset in Sec.~\ref{sec:model} with composite states.
Under the paradigm of partial compositeness~\cite{Kaplan:1991dc}, composite partners $\Psi_{L,R}$ and their counterparts $\widetilde{\Psi}_{L,R}$ in the color-neutral sector are introduced to mimic the Kaluza-Klein states in the holographic setup. The fermion assignment is shown as the moose diagram in Fig.~\ref{fig4}.
The link field $U$ is defined as, in the unitary gauge,
\bea
U=\left(
\baa{ccc}
\bold1_{3\times 3}&\ &\ \\
\ &c_h&s_h\\
\ &-s_h&c_h\\  
\eaa
\right),
\eea
which corresponds to the Wilson line along the fifth dimension in the $5$D model.

With the $SO(4)_2$ gauge symmetry on the $2$-site, we can decompose the composite $\Psi$ ($\widetilde{\Psi}$) to $\Psi^{(4)}$ ($\widetilde{\Psi}^{(4)}$) and $\Psi^{(1)}$ ($\widetilde{\Psi}^{(1)}$).
The fermionic Lagrangian reads
\bea
\begin{aligned}
\mathcal{L}=\ &yf\bar{Q}_LU\Psi_R-M\bar{\Psi}_{L}\Psi_{R}-m\bar{\Psi}^{(1)}_{L}t_R\\
&+\widetilde{y}f\bar{\widetilde{Q}}_LU\widetilde{\Psi}_R-\widetilde{M}\bar{\widetilde{\Psi}}_{L}\widetilde{\Psi}_{R}-\widetilde{m}\bar{\widetilde{\Psi}}^{(1)}_{L}\widetilde{T}_R\\
&-\widetilde{m}_{q}\bar{\widetilde{Q}}_{L}\widetilde{Q}_{R}+\text{h.c.}\ ,\\
\end{aligned}
\label{composite}
\eea
where mass splittings of $\Psi_{4,1}$ ($\widetilde{\Psi}_{4,1}$) are assumed to be zero. 
Since $Q_L$ and $\widetilde{Q}_{L,R}$ form incomplete $SO(5)$ multiplets, $SO(5)_1$ is explicitly broken. Without introducing mass terms $m\bar{\Psi}^{(1)}_{L}t_R$ and $\widetilde{m}\bar{\widetilde{\Psi}}^{(1)}_{L}\widetilde{T}_R$, the $SO(5)_2$ remains unbroken, then the Higgs is an exact Goldstone. The non-vanishing Higgs potential can only exist when both $SO(5)_1$ and $SO(5)_2$ are explicitly broken, which is referred as collective symmetry breaking~\cite{ArkaniHamed:2002qy,Foadi:2010bu}. Therefore, the PNGB Higgs is doubly protected by the collective symmetry and the $SU(6)_c$ or $Z_2$ symmetry between the top quark and the hidden top partners.

In both frameworks, considering low energy effective theory, one can integrate out the bulk dynamics or the composite fermions $\Psi$ and $\widetilde{\Psi}$ in Eq.~\ref{composite} and match to the following effective Lagrangian
\bea
\begin{aligned}
\mathcal{L}_{\text{eff}}=\ &\bar{t}_L p\!\!\!/\Pi_{t_L}t_L+\bar{t}_R p\!\!\!/\Pi_{t_R}t_R-\left(\bar{t}_L\Pi_{t_Lt_R}t_R+\textnormal{h.c.}\right)\\
&+\bar{\widetilde{L}} p\!\!\!/\widetilde{\Pi}_{L}\widetilde{L}+\bar{\widetilde{R}} p\!\!\!/\widetilde{\Pi}_{R}\widetilde{R}-\left(\bar{\widetilde{L}}\widetilde{\Pi}_{LR}\widetilde{R}+\textnormal{h.c.}\right)\ ,
\end{aligned}
\label{holo}
\eea
where the hidden top partners are denoted as
\bea
\widetilde{L}=
\left(
\baa{c}
\widetilde{t}_L\\ \widetilde{T}_L
\eaa
\right),\ \ 
\widetilde{R}=
\left(
\baa{c}
\widetilde{t}_R\\ \widetilde{T}_R
\eaa
\right).\ \ \ 
\eea
Following the holographic approach~\cite{Contino:2004vy,Contino:2003ve,Agashe:2004rs,Serone:2009kf}, the Higgs potential is derived as
\begin{align}
&V(h)=-\frac{2N_c}{16\pi^2}\int \textnormal{d}Q^2 Q^2\ \textnormal{log}\left[\Pi_{t_L}\Pi_{t_R}\cdot Q^2+|\Pi_{t_Lt_R}|^2\right]\nn\\
&-\frac{2\widetilde{N}_c}{16\pi^2}\int \textnormal{d}Q^2 Q^2\ \textnormal{Tr}\left\{\textnormal{log}\left(1+\frac{\widetilde{\Pi}_{LR}\widetilde{\Pi}^{-1}_{R}\widetilde{\Pi}^\dagger_{LR}\widetilde{\Pi}^{-1}_{L}}{Q^2}\right)\right.\nn\\
&+\left.\textnormal{log}\left(1+(\widetilde{\Pi}_{L}-\widetilde{\Pi}_{L0})\widetilde{\Pi}^{-1}_{L0}\right)+\textnormal{log}\left(1+(\widetilde{\Pi}_{R}-\widetilde{\Pi}_{R0})\widetilde{\Pi}^{-1}_{R0}\right)\right\},
\label{potential_composite}
\end{align}
where the loop momentum is defined in Euclidean space, i.e., $Q^2=-p^2$, and $\widetilde{\Pi}_{L0, R0}$ denote the Higgs-independent part of the wave functions defined as Eq.~\ref{holo}.

After tedious calculations, we obtain the finite Higgs potential without $\log$-dependence
\begin{align}
V(h)\simeq&-\frac{3\left(a\ y^2f^2m^2-\widetilde{a}\ \widetilde{y}^2f^2\widetilde{m}^2\right)}{16\pi^2}s_h^2\nn\\
&\ \ \ +\frac{3\left(b\ y^4f^4+\widetilde{b}\ \widetilde{y}^4f^4\right)}{16\pi^2}s_h^4\ ,
\label{potential_simplify}
\end{align}
where parameters $a,\widetilde{a},b,\widetilde{b}$ depend on the bulk masses and mixings in the holographic model, or the parameters in Eq.~\ref{composite} for the composite model. The $s_h^2$ term above contains two opposite contributions from the SM and the hidden sector, which suppress the relative magnitude compared to the $s_h^4$. This realizes vacuum misalignment.
 The corresponding fine-tuning level of our model is similar to the one of other composite Higgs models~\cite{Panico:2012uw,Bellazzini:2014yua}.

\section{Model Implications and Conclusions}

This MNNM setup solves the little hierarchy problem with minimal hidden sector. The hidden sector contains dark $SU(3)^\prime_c$ color, under which new electroweak singlet and doublet fermions are charged. These fermions are responsible for generating the radiative Higgs potential with their masses around scale $f$. Since the dark color group is confined at around GeV scale and no new fermions are lighter than that, the heavy fermions exhibit quirk behavior~\cite{Burdman:2008ek,Cai:2008au} and form macroscopic bound states~\cite{Kang:2008ea}. Depending on the electroweak charges of these fermions, the top partners can be neutral or charged particles, i.e., $Q=0,2/3,1$ etc. Discovering these exotic bound states forming from quirks is one of the smoking-gun signatures at colliders. 
For charged top partners, they can be produced through the Drell-Yan process and then form bound states~\cite{Harnik:2011mv,Chacko:2015fbc,Knapen:2017kly,Farina:2017cts}.
For neutral top partners, it is more promising to detect them at the Large Hadron Collider (LHC) with the possible displaced vertex signature~\cite{Craig:2015pha,Curtin:2018mvb} from Higgs exotic decay.

\begin{figure}[!h]
\includegraphics[scale=0.5]{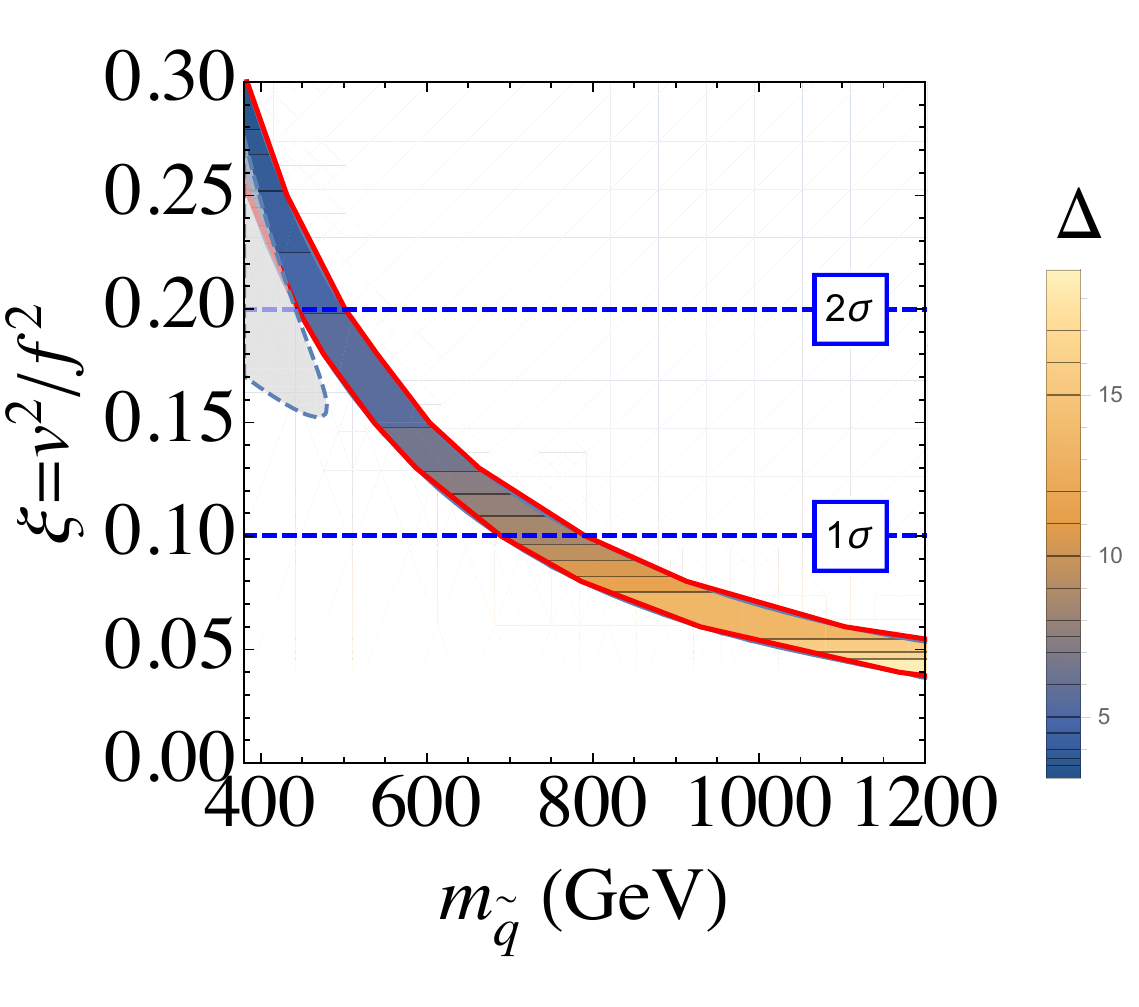}
\caption{The colored region denotes the viable parameter space on ($\xi$, $m_{\widetilde{q}}$), in which vacuum misalignment and the $125$ GeV Higgs mass are obtained, with electric charge of the top partners unspecified. The color bar on the right shows the fine-tuning level. The blue dashed lines denote the $1\sigma$, $2\sigma$ contours allowed by Higgs data, assuming electric charge $Q=1$ for the hidden tops. The gray shaded region denotes the T parameter constraints to the top partners.}
\label{parameter}
\end{figure}

Current searches at the LHC put constraints on the model parameters $f$ and $m_{\widetilde{q}}$, which determines the low energy spectrum and couplings in MNNM. Since the Higgs boson is a PNGB connecting the SM and the hidden sector, the tightest constraint is from Higgs coupling measurements~\cite{Aad:2015zhl,Khachatryan:2016vau,ATLAS:2018doi}. We perform a global analysis on the Higgs nonlinearity parameter $\xi=v^2/f^2$ and new fermion mass parameter $m_{\widetilde{q}}$ using latest Higgs data encoded in the program Lilith~\cite{Bernon:2015hsa}. 
As shown in Fig.~\ref{parameter}, the blue dashed lines read $\xi<0.1\ (0.2)$ at $1\sigma\ (2\sigma)$ confidence level. 
We also consider constraints on vector-like top partners from electroweak precision tests, using the oblique parameters S, T~\cite{Peskin:1991sw,Baak:2012kk}. The gray shaded region in Fig.~\ref{parameter} shows the dominated constraint from the T parameter on singlet and doublet fermion masses and mixing~\cite{Lavoura:1992np,Cynolter:2008ea}.
To obtain the correct vacuum misalignment and the $125$ GeV Higgs mass, model parameters need to be within the colored region in Fig.~\ref{parameter}, with $\Lambda\subset[3\ \text{TeV}, 10\ \text{TeV}]$ and $y\subset[0.86, 0.98]$ corresponding to the running top quark mass at TeV scale. The strong correlation between $\xi$ and $m_{\widetilde{q}}$ shows the cancellation of singlet and doublet top partner's contributions on the Higgs quadratic term, and thus determines the fine-tuning level of the model. 
According to Eq.~\ref{gen_po}, we define the fine-tuning level~\cite{Bellazzini:2014yua} as
\bea
\Delta\equiv \frac{\gamma\mid_{m_{\widetilde{q}}= 0}}{\gamma_0} \cdot \frac{\beta\mid_{m_{\widetilde{q}}= 0}}{\beta_0}\sim \mathcal{O}(10)\ ,
\eea
where $\gamma_0$ and $\beta_0$ are the correct values for obtaining vacuum misalignment and the Higgs mass.
As shown in Fig.~\ref{parameter}, the smaller $\xi$, the larger $m_{\widetilde{q}}$, and severer fine-tuning. The SM is recovered as $m_{\widetilde{q}}\to\infty$ and $\xi\to 0$.

The hidden sector can naturally accommodate dark matter candidate. If the top partners are charge neutral, they form the lightest dark baryon which serves as dark matter in the asymmetric dark matter scenario~\cite{Kaplan:2009ag,Zurek:2013wia,Kribs:2009fy,Farina:2015uea,Garcia:2015toa}. In this minimal setup, it is not easy to identify the dark baryon or meson as WIMP dark matter because of strong interactions in the dark color sector. One needs to introduce either light dark-colored fermions to have SIMP dark matter~\cite{Hochberg:2014dra} or leptons in the hidden sector to have singlet-doublet fermion dark matter~\cite{Yaguna:2015mva,Garcia:2015loa}. More detailed study on dark matter phenomenology is beyond the scope of this paper.

Overall, we lay out the basic setup of a minimal neutral naturalness model and its UV extension with emphasis on generating the realistic Higgs potential. This setup contains very rich phenomenology which cannot be expanded in this paper, such as dark hadron spectra, collider signatures, cosmological implications, heavy composite particles, etc. To distinguish this model from other neutral naturalness models, we need to know the mass relation of the doublet and singlet fermions after their discovery. In future, more efforts are needed on exploring this neutral naturalness scenario.

\section*{Acknowledgments} 
JHY is supported by the National Science Foundation of China under Grants No. 11875003 and the Chinese Academy of Sciences (CAS) Hundred-Talent Program. LXX and SHZ are supported in part by the National Science Foundation of China under Grants No. 11635001, 11875072.

\section*{Supplementary Materials}

In this section, we present details on the specific forms of the form factors defined in the general effective Lagrangian of Eq.~\ref{holo}. Then we briefly list the numerical results in the composite model and the $5$D holographic model. 

In composite MNNM, we have the following form factors in the SM sector and the hidden sector after integrating out the composite fermions,
\begin{align}
&\Pi_{t_L}=1-\frac{y^2f^2}{2}\frac{1}{p^2-M^2},\ \ \Pi_{t_R}=1-\frac{m^2}{p^2-M^2},\nn\\
&\Pi_{t_Lt_R}=\frac{iyf}{\sqrt{2}}s_h\frac{Mm}{p^2-M^2}, 
\label{ffsm}
\end{align}
\begin{align}
&\widetilde{\Pi}_L=
\left(
\baa{cc}
1-\frac{\widetilde{y}^2f^2}{2(p^2-\widetilde{M}^2)}& 0\\ 
0 & 1-\frac{\widetilde{y}^2f^2}{p^2-\widetilde{M}^2}
\eaa
\right),\ \nn\\
&\widetilde{\Pi}_R=
\left(
\baa{cc}
1& 0\\ 
0 & 1-\frac{\widetilde{m}^2}{p^2-\widetilde{M}^2}
\eaa
\right),\ \nn\\
&\widetilde{\Pi}_{LR}=
\left(
\baa{cc}
\widetilde{m}_{q}& \frac{-i\widetilde{y}f}{\sqrt{2}}s_h\frac{\widetilde{m}\widetilde{M}}{p^2-\widetilde{M}^2}\\ 
0 & \widetilde{y}fc_h\frac{\widetilde{m}\widetilde{M}}{p^2-\widetilde{M}^2}
\eaa
\right).
\label{ff}
\end{align}
We see explicitly the wave functions do not depend on Higgs, while $\Pi_{t_Lt_R}$ and $\widetilde{\Pi}_{LR}$ have Higgs filed dependence. 
\begin{figure}
\includegraphics[scale=0.2]{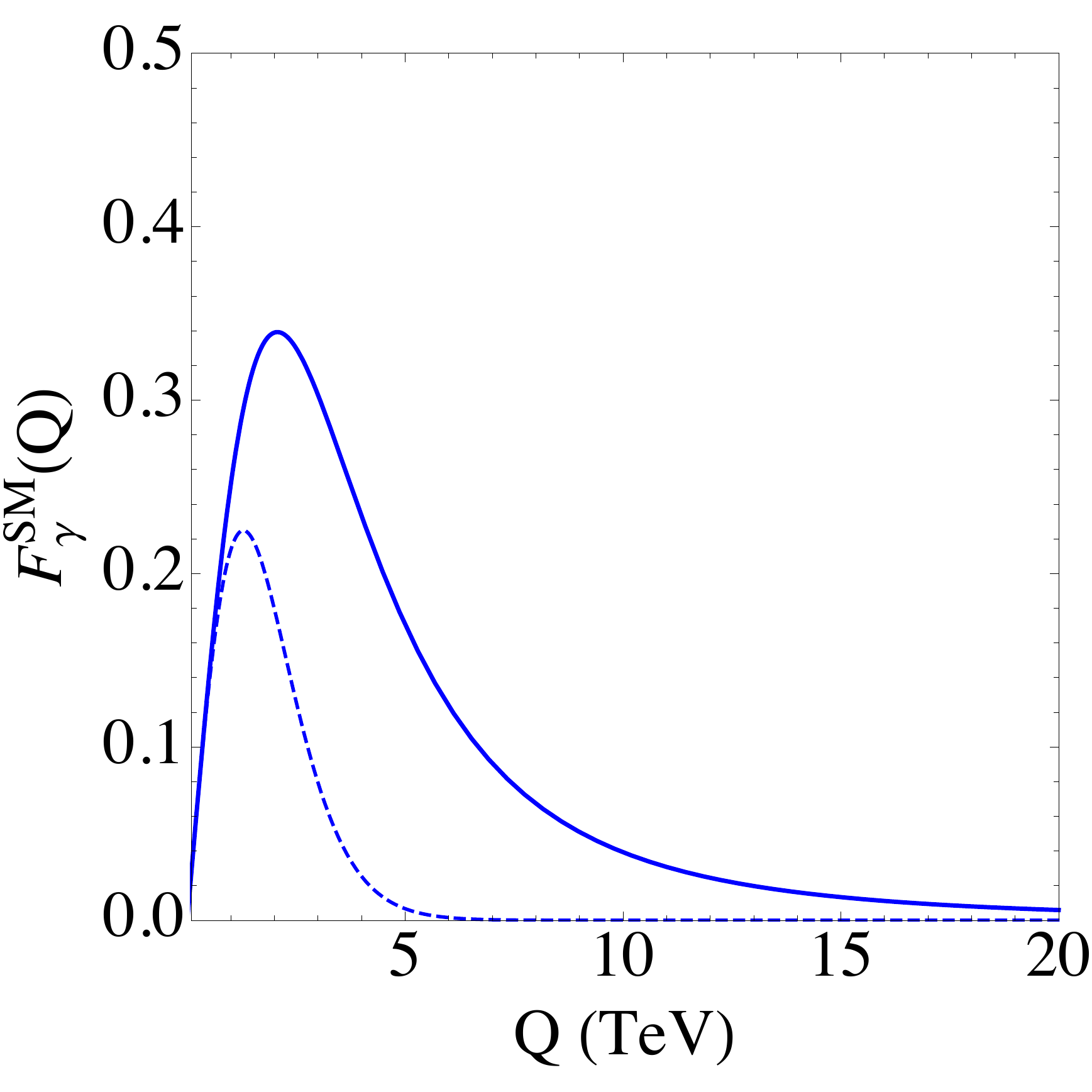}
\includegraphics[scale=0.2]{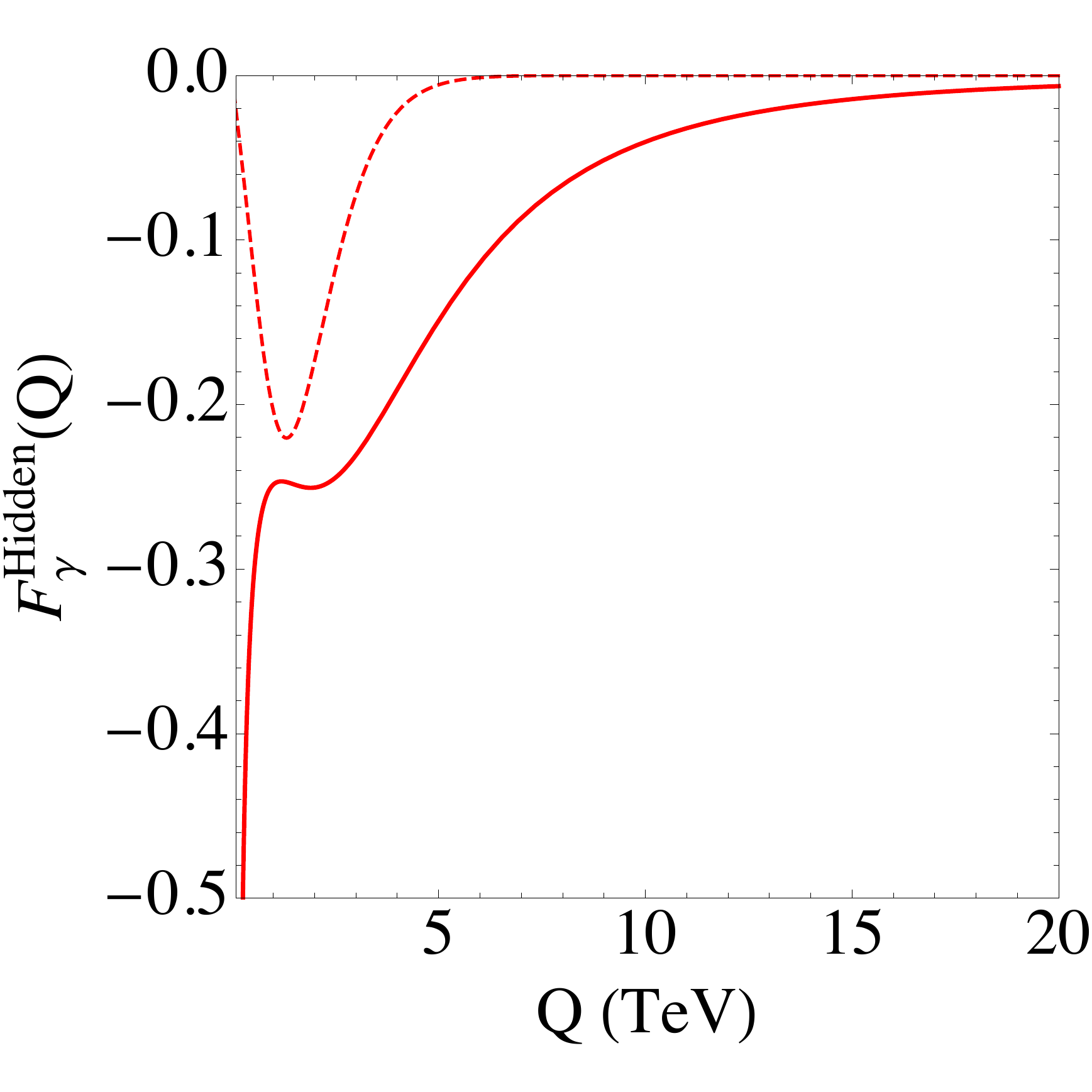}
\includegraphics[scale=0.2]{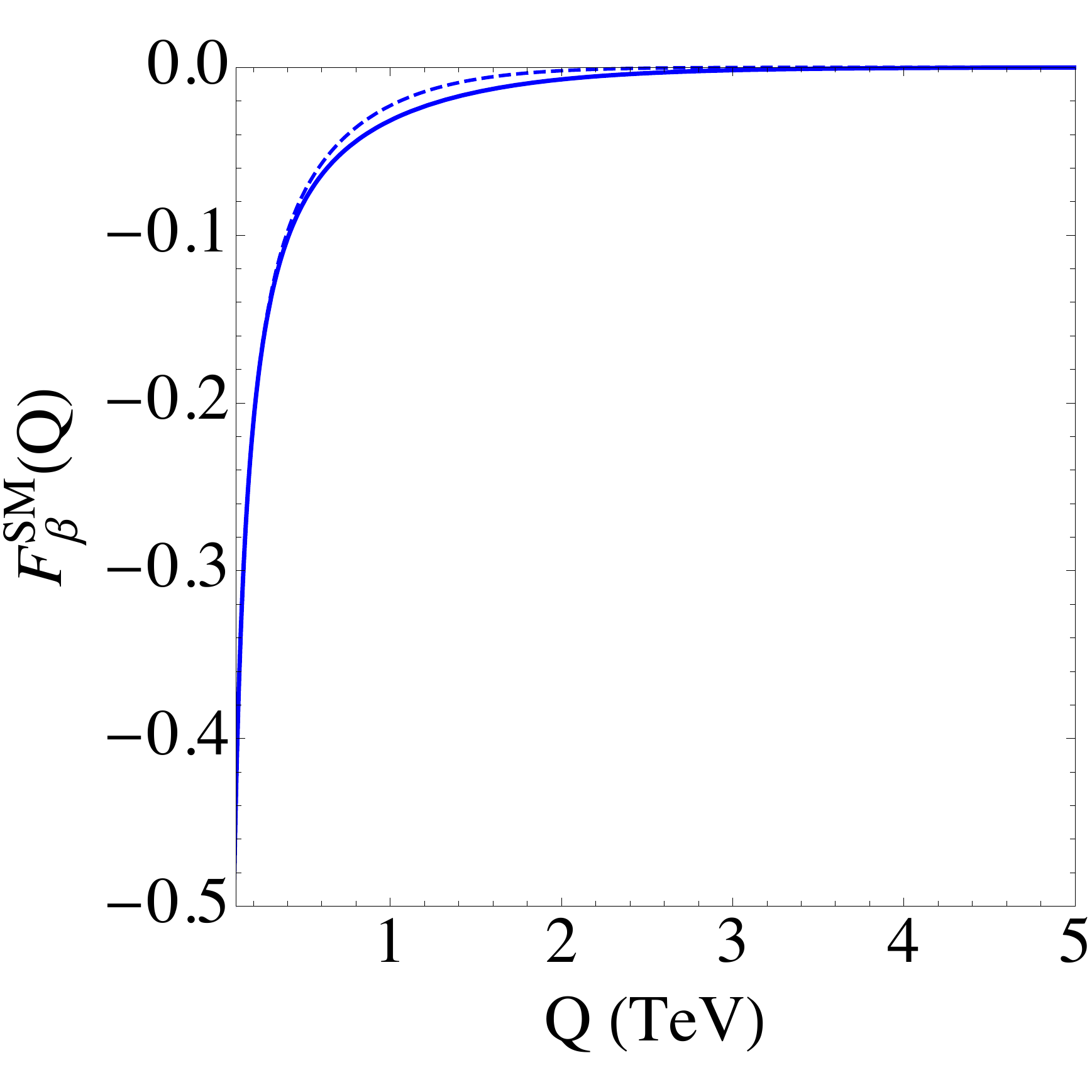}
\includegraphics[scale=0.2]{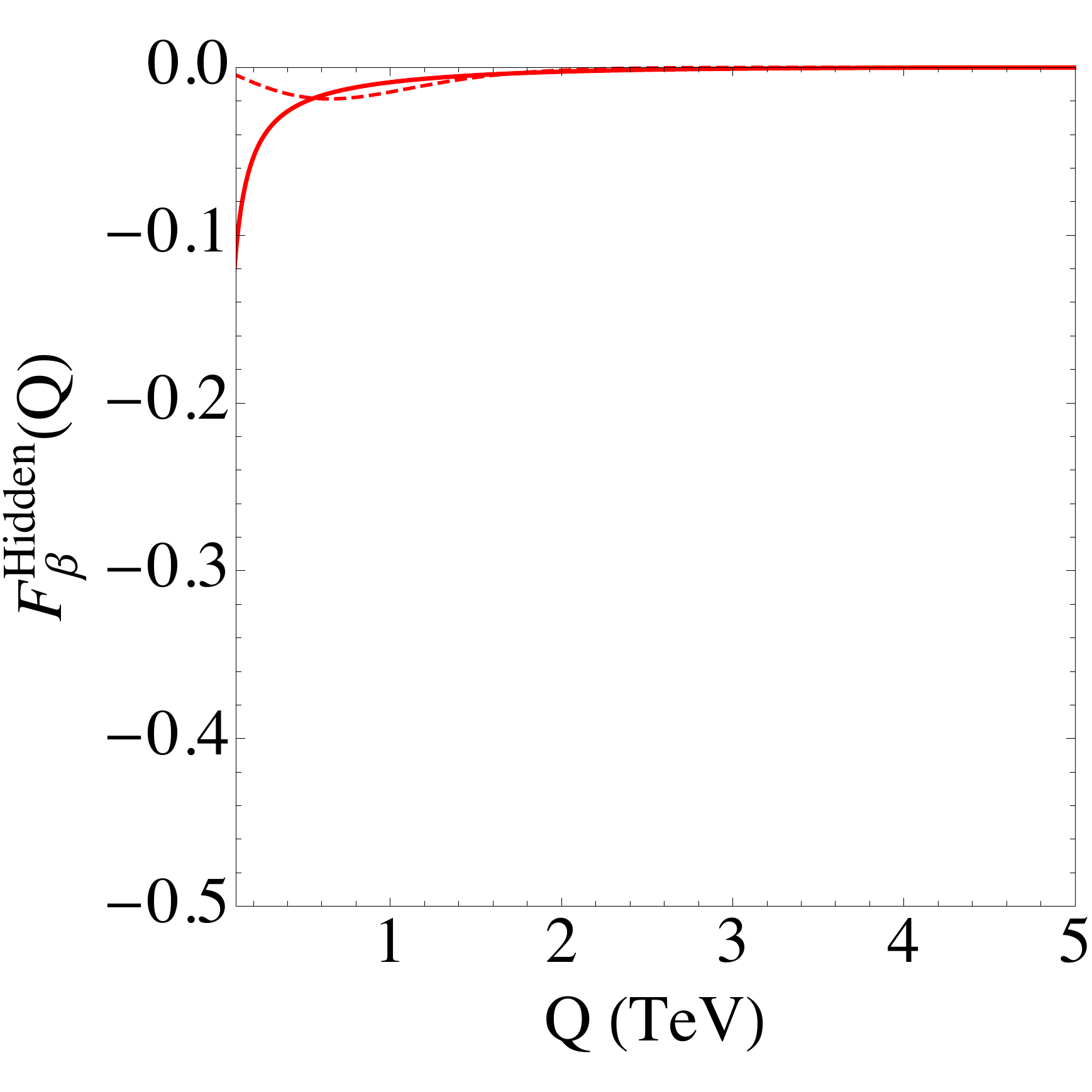}
\caption{The integrands of the Higgs potential, combinations of the form factors, for the composite neutral model (solid lines) and the $5$D model (dashed lines). The blue lines denote the contribution of the SM top quark to $\gamma$ (top left) and $\beta$ (bottom left), while the red lines denote the contribution of the neutral tops to $\gamma$ (top right) and $\beta$ (bottom right).}
\label{composite_FF}
\end{figure}

In the $5$D holographic model, we have the form factors as follows
\begin{align}
&\Pi_{t_L}=\frac{\Pi_{L}(0)}{p}+\frac{\Pi_{L}(m)-\Pi_{L}(0)}{2\ p}s_h^2,\nn\\
&\Pi_{t_R}=\frac{\Pi_{R}(m)}{p},\nn\\
&\Pi_{t_Lt_R}=\frac{i\ \Pi_{LR}(m)}{\sqrt{2}}s_h, 
\end{align}
\begin{align}
&\widetilde{\Pi}_L=
\left(
\baa{cc}
\frac{\widetilde{\Pi}_L(0)}{p}+\frac{\widetilde{\Pi}_L(\widetilde{m})-\widetilde{\Pi}_L(0)}{2\ p}s_h^2 & i\frac{\widetilde{\Pi}_L(\widetilde{m})-\widetilde{\Pi}_L(0)}{\sqrt{2}\ p}s_hc_h\\ 
-i\frac{\widetilde{\Pi}_L(\widetilde{m})-\widetilde{\Pi}_L(0)}{\sqrt{2}\ p}s_hc_h & \frac{\widetilde{\Pi}_L(\widetilde{m})}{p}-\frac{\widetilde{\Pi}_L(\widetilde{m})-\widetilde{\Pi}_L(0)}{p}s_h^2
\eaa
\right),\ \nn\\
&\widetilde{\Pi}_R=
\left(
\baa{cc}
1& 0\\ 
0 & \frac{\widetilde{\Pi}_R(\widetilde{m})}{p}
\eaa
\right),\ \nn\\
&\widetilde{\Pi}_{LR}=
\left(
\baa{cc}
\widetilde{m}_{q}& -\frac{i}{\sqrt{2}}\widetilde{\Pi}_{LR}(\widetilde{m}) s_h\\ 
0 & -\widetilde{\Pi}_{LR}(\widetilde{m}) c_h
\eaa
\right), 
\label{ff}
\end{align}
where $\Pi_{L,R,LR}(m)$ and $\widetilde{\Pi}_{L,R,LR}(\widetilde{m})$ are the form factors with non-zero mixing parameters on the IR brane, while $\Pi_{L,R,LR}(0)$ and $\widetilde{\Pi}_{L,R,LR}(0)$ are the form factors with vanishing mixing parameters. We see the Higgs field dependence vanishes when $m\to 0$ and $\widetilde{m}\to 0$.
The concrete expressions of the above form factors are too lengthy, we will present them in a forthcoming publication.

With above results, we see the Higgs dependence in the composite model and the $5$D model is similar to each other, despite the fact that $\Pi_{t_L,t_R}$ and $\widetilde{\Pi}_{L,R}$ in the $5$D model depend on the Higgs field. 
The Higgs potential is obtained as 
\begin{align}
\gamma&=\frac{3}{4 \pi^2}\int \text{d}Q \ (F^{\text{SM}}_\gamma+F^{\text{Hidden}}_\gamma)\ ,\\
\beta&=-\frac{3}{4 \pi^2}\int \text{d}Q \ (F^{\text{SM}}_\beta+F^{\text{Hidden}}_\beta)\ ; 
\end{align}
see Eq.~\ref{gen_po} for the definitions of $\gamma$ and $\beta$. Here $F^{\text{SM}}_{\gamma\ (\beta)}$ and $F^{\text{Hidden}}_{\gamma\ (\beta)}$ represent form factor combinations.
In calculations, we only include the leading term in the expansion of $\textnormal{log}\left(1+\widetilde{\Pi}_{LR}\widetilde{\Pi}^{-1}_{R}\widetilde{\Pi}^\dagger_{LR}\widetilde{\Pi}^{-1}_{L}/Q^2\right)$ in the $5$D model, while including the leading two terms in the composite model. 
Note that an IR cutoff is needed in the evaluation of the Higgs potential, and we choose its value as $100$ GeV.
Fig.~\ref{composite_FF} shows the form factor combinations, which are the integrands of the Higgs potential according to Eq.~\ref{potential_composite}, in the composite model and the $5$D model respectively.
We see explicitly the cancellation between $F^{\text{SM}}_\gamma$ and $F^{\text{Hidden}}_\gamma$ in $\gamma$, and all the form factors vanishes when $Q\to \infty$, thus the Higgs potential is finite.

\bibliographystyle{utphys}
\bibliography{neutraltop}

\end{document}